\documentstyle[12pt,a4]{article}

\newcommand{\be}{\begin{equation}}
\newcommand{\ee}{\end{equation}}
\newcommand{\bea}{\begin{eqnarray}}
\newcommand{\eea}{\end{eqnarray}}
\newcommand{\beast}{\begin{eqnarray*}}
\newcommand{\eeast}{\end{eqnarray*}}

\newcommand{\bfm}{{\bf M}}
\newcommand{\bfphi}{{\bf \Phi}}
\newcommand{\phidagphi}{\bfphi^\dagger\bfphi}

\begin{document}
\begin{titlepage}
\begin{flushright}
DO-TH-95/04\\
March 1995
\end{flushright}

\vspace{20mm}
\begin{center}
{\Large \bf
Fluctuation corrections to bubble nucleation}

\vspace{10mm}

{\large  J. Baacke\footnote{e-mail:~
baacke@het.physik.uni-dortmund.de}} \\
\vspace{15mm}

{\large Institut f\"ur Physik, Universit\"at Dortmund} \\
{\large D - 44221 Dortmund , Germany}
\vspace{25mm}

\bf{Abstract}
\end{center}
The fluctuation determinant which determines the
preexponential factor of the transition rate
for minimal bubbles is computed for the electroweak
theory with $\sin \Theta_W = 0$. As the basic action
we use the three-dimensional high-temperature action
including, besides temperature dependent masses,
the $T \Phi^3$ one-loop contribution whcih makes the phase
transition first order. The results show that this term
(which has then to be subtracted from the exact result)
gives the dominant contribution to the one-loop effective
action. The remaining correction is of the order of, but
in general larger than the critical bubble action.
The results for the Higgs field fluctuations are compared
with those of an approximate
heat kernel computation of Kripfganz et al., good
agreement is found for small bubbles, strong deviations
for large thin-wall bubbles.

\end{titlepage}


\section{INTRODUCTION}
\setcounter{equation}{0}
The electroweak phase transition is at present the
object of extensive investigations \cite{Sin}. If the
phase transition is first order, which is possibly the case
if the mass of the Higgs boson is not too large,
the phase transition occurs via bubble nucleation. Bubble nucleation
can have various consequences for cosmology in the early universe.
The possibility of baryogenesis in bubble walls has been investigated
recently by many authors
(see e.g. \cite{KuRuSha,CoKaNe}), reheating
after the phase transition could be
mediated by bubble nucleation and subsequent coalescence, the creation
of inhomogeneities by bubble formation
could be observable (see e. g. \cite{TuWeWi,LiLeTu,DLHLL} for
representative discussions of the physics of bubble nucleation and
growth).

Bubble nucleation is described usually within the
reaction rate theory formulation
of Langer \cite{La} or, equivalently, the semiclassical
approach to quantum field theory by
Coleman and Callan \cite{Co,CaCo}.
This formulation requires the existence
of a saddle point in configuration space, the minimal bubble,
with one unstable mode, possible zero modes and real frequency
fluctuation modes. The leading term in the tunnelling rate is given
by the negative exponential of the minimal bubble action,
the corrections arise from integrating
out the fluctuations in the Gaussian approximation, leading to
a fluctuation determinant prefactor whose negative logarithm
is the 1-loop effective action. If the leading approximation is
good this prefactor should be of order $1$, substantial prefactors
have however been found in the case of the sphaleron transition,
both from bosonic \cite{CarLi,BaaJu,DyaGoe1} as also fermionic
\cite{DyaGoe2} fluctuations. It is therefore of interest
to investigate how strongly these prefactors modify the
leading order approximation to the bubble nucleation rate.

Here we present an exact computation
of the bosonic fluctuation determinant
of the critical bubble. As the basic action is determined by
the usual Higgs potential with just one minimum at the
classical expectation value, some fluctuation effects
have to be included already at the tree level in order to
allow for minimal bubble solutions. The exact fluctuation
determinant should then reproduce those in order to justify
this modification of the leading order action.
Following the basic work of Coleman and Weinberg \cite{CoWe}
such  modified
actions have been proposed by many authors
\cite{DLHLL,KiLi,Sha,AnHa} and used to describe
the bubble nucleation in leading order. To be specific
we use here the one given by Dine et al.\cite{DLHLL}
which was also the basis of a recent approximate computation
of the 1-loop Higgs fluctuations by Kripfganz et al.
\cite{KriLaSch}.

The plan of this paper is as follows:
In the next section we will introduce the model and set up
the basic relations for the bubble nucleation rate.
In section 3 we will discuss the structure of the
fluctuation operator, in particular its
partial wave decomposition.
The computation of its determinant, based on a very useful
theorem, will be described in section 4.
In the final section we will present some
results and conclusions.

\section{Basic relations}
\setcounter{equation}{0}

The three-dimensional high-temperature action
is given, in the formulation by Dine et al. \cite{DLHLL},
by
\bea \label{htac}
S_{ht} & =&\frac{1}{g_3(T)^2}
\int d^3x \left[\frac{1}{4}F_{ij}F_{ij}+
\frac{1}{2}(D_i\Phi)^\dagger (D_i \Phi)
 + V_{ht}(\phidagphi) \right. \nonumber \\
&&\left.+ \frac{1}{2} A_0 \left(-D_iD_i +\frac{1}{4} \phidagphi
\right )A_0 \right] \; .
\eea
Here the coordinates and fields have been rescaled as \cite{CarMcL}
\be \label{scaling1}
\vec x \to \frac{\vec x}{g v(T)}, \;
\bfphi \to v(T) \bfphi, \; A \to v(T) A \; .
\ee
The vacuum expectation value  $v(T)$ is defined as
\be \label{scale1}
v^2(T)=\frac{2D}{\lambda_T} (T_0^2 - T^2)\;.
\ee
$T_0$ is the temperature at which the high-temperature
potential $V_{ht}$ changes its extremum at
$\bfphi = 0$ from a minimum at $T > T_0$ to a maximum at
$T < T_0$. The temperature dependent coupling of the
three-dimensional theory is defined as
\be \label{htcoupl}
g_3(T)^2 = \frac{gT}{v(T)} \; .
\ee
In terms of the zero temperature parameters
we have $m_W = g v_0/2$, $m_H = \sqrt{2\lambda} v_0$ with
$v_0 = 246 $ GeV and we use the definitions of Dine et al.
\cite{DLHLL} modified by setting $\Theta_W=0$ and therefore
$m_W=m_Z$ :
\bea \label{coeffs}
D&=& (3m_W^2+2m_t^2)/8v_0^2 \nonumber \\
E&=& 3 g^3/32 \pi \nonumber \\
B&=& 3 ( 3m_W^4 - 4 m_t^4)/64\pi^2v_0^4 \nonumber\\
T_0^2&=& (m_H^4-8 v_0^2 B)/4D  \\
\lambda_T&=& \lambda -3(3m_W^4\ln\frac{m_w^2}{a_BT^2}
-4 m_t^4 \ln\frac{m_t^2}{a_F T^2})/16 \pi^2 v_0^4 \; .
\eea
In terms of these parameters the high-temperature
potential is given by
\be \label{htpot1}
V_{ht}(\phidagphi) = \frac{\lambda_T}{4g^2}
\left( (\phidagphi)^2- 2\phidagphi  -\frac{4 E}{\lambda_T v(T)}
(\phidagphi)^{3/2} \right) \; .
\ee

The rescaling Eq. (\ref{scaling1}) with the
scale $v(T)$ makes sense only for $T < T_0$. On the other hand
the high-temperature potential has, before rescaling, a
secondary minimum at $|\bfphi| = \tilde v (T)$ with
\be
\tilde v (T) = \frac{3 E T}{2\lambda}+
\sqrt{\left(\frac{3ET}{2\lambda}\right)^2+v^2(T)} \; .
\ee
This minimum is degenerate with the one at $\bfphi =0$
at a temperature defined implicitly by
\be
 T_C = T_0/\sqrt{1 - E^2/\lambda_{T_C}} \; .
\ee
$T_C$ marks the onset of bubble formation by
thermal barrier transition.
In the work of Hellmund et al. \cite{HeKriSch} and
Kripfganz et al. \cite{KriLaSch} the vacuum expectation
value of the broken symmetry phase $\tilde v (T)$ is chosen
for the rescaling of the fields, i.e. in Eq. (\ref{scaling1})
$v(T)$ is replaced by $\tilde v(T)$, the high-temperature
coupling constant
Eq.(\ref{htcoupl}) is redefined analoguously and
denoted\footnote{
Our notation differs from the one of Refs. \cite{KriLaSch,HeKriSch}.}
as $\tilde g_3 (T)$.
By this change of scale the high-temperature potential changes
as well; it becomes \footnote{We do not introduce a tilde for
the rescaled fields.} \cite{HeKriSch}
\be \label{htpot2}
V_{ht}(\phidagphi) =
\frac{\lambda_T}{4g^2} \left( (\phidagphi)^2-
\epsilon(T) (\phidagphi)^{3/2} + (\frac{3}{2}
\epsilon(T)-2)\phidagphi \right )
\ee
with
\be \label{epsdef}
\epsilon(T)=\frac{4}{3}\left ( 1 - \frac{v(T)^2}{\tilde v (T)^2}
\right ) \; .
\ee

The action and its rescaling differ slightly from that of
Hellmund et al. \cite{HeKriSch} and of Kripfganz et al. \cite{KriLaSch}.
In contrast to the former we do not mimic the influence of a Debye
mass by decoupling the longitudinal degrees of freedom.
In contrast to the second one we include only the $\Phi^3$
contribution of
gauge field and would-be Goldstone degrees of freedom as in
Ref. \cite{Sha}. This form of the $\Phi^3$ contribution was
found to yield a good approximation for the exact results
in the case of the sphaleron \cite{CarLi,BaaJu}, at least in the case
$m_H/m_W \ll 1$. We will find, indeed, that this term dominates
the effective action.

The process of bubble nucleation is - within the approach
of Langer \cite{La} and Coleman and Callan
\cite{Co,CaCo}, followed by the work of
Affleck \cite{Af}, Linde \cite{Li} and others
 - described by the rate
\be \label{rate}
\Gamma/V = \frac{\omega_-}{2 \pi} \left (
\frac{\tilde S}{2\pi}\right )^{3/2}\exp(-\tilde S)~ {\cal J}^{-1/2}
\; . \ee
Here $\tilde S$ is the high-temperature action, Eq. (\ref{htac}),
with the new rescaling, minimized by a classical minimal
bubble configuration (see below), $\cal J$ is the fluctuation
determinant which describes the next-to-leading part of the
semiclassical approach and which will be defined below;
its logarithm is related to the 1-loop
effective action by
\be
S^{1-l}_{eff} = \frac{1}{2} \ln {\cal J} \; .
\ee
Finally $\omega_-$ is the absolute value of
the unstable mode frequency.

The classical bubble configuration is described by
a vanishing gauge field and a real spherically symmetric
Higgs field
$\Phi(r) = |\bfphi| (r)$ which is a solution of the
Euler-Lagrange equation
\be \label{Clbub}
-\Phi''(r)-\frac{2}{r}\Phi'(r)+\frac{d V_{ht}}{d\Phi(r)} = 0
\ee
with the boundary conditions
\be
\lim_{r\to\infty}
\Phi(r)=0 ~~
 {\rm and} ~~ \Phi'(0)=0 \; .
\ee
This differential equation can be
solved numerically e.g. by the shooting method. The solution
will be denoted as $H_0(r)$.

\section{Fluctuation analysis}

In terms of the action $S$ the fluctuation operator is defined
generally as
\be \label{flucop}
{\cal M}_{ab} = \frac{\delta^2 S}{\delta \phi_a \delta \phi_b}
|_{\phi=\phi_{bubble}},
\ee
where $\phi_a$ stands for the various gauge and Higgs field
components and $\phi_{bubble}$ is the field configuration
of the minimal bubble.
 An analoguous derivative, taken at $\phi = \phi_{vac} \equiv 0$
defines the vacuum fluctuation operator ${\cal M}^0$.
In both configurations the gauge fields vanish, the Higgs field is
given by
\be \label{higbub}
\bfphi = H_0(r) \left( \begin{array}{c} 0 \\ 1 \end{array}
\right)
\ee
in the bubble configuration and vanishes in the vacuum.

The fluctuation determinant $\cal J$ appearing in the rate formula
is defined by
\be
{\cal J} =\frac{\det''{\cal M}}{\det {\cal M}^0} \; .
\ee
Here the symbol $\det''$ denotes the determinant with removed
translation zero modes and with the unstable mode frequency
replaced by its absolute value.

The analysis of fluctuations of the minimal bubble
can be related to a similar analysis
performed recently for the electroweak sphaleron
without gauge fixing in Ref. \cite{BaaLa} and in the
't Hooft-Feynman background gauge in Ref. \cite{BaaJu}.
We will use this latter analysis. One can take over the fluctuation
operator with two modifications which represent
at the same time essential simplifications:
\\ \noindent
- the high-temperature effective potential
has to be modified from the one in Eq. (\ref{htpot1}) to
the one in Eq. (\ref{htpot2});
\\ \noindent
- the sphaleron and the broken symmetry vacuum
configurations are replaced by the bubble and
the symmetric vacuum configurations defined above.

Furthermore we use here (see Eq. (\ref{scaling1})) for the coordinates
the scale $(g \tilde v)^{-1}$ instead of the scale
$M^{-1}_W= 2/gv$ used in Ref. \cite{BaaJu}.

The expansion of gauge and Higgs fields reads then \cite{BaaJu}
\bea
W_\mu^a & = & a_\mu^a \nonumber \\
\bfphi & = & (H_0 + h + \tau^a \phi^a)\left( \begin{array}{c}
 0 \\ 1 \end{array} \right) \; .
\eea
Here the fields denoted with small letters , $ a_\mu^a, h $
 and $ \phi_a$ are the fluctuating fields.

Before we discuss fluctuations we have to fix the gauge.
We work here in the 't Hooft-Feynman background gauge.
The gauge conditions read
\be \label{backgauge}
{\cal F}_a = \partial_\mu a^\mu_a+ \frac{1}{2}H_0 \phi_a =0 \; .
\ee
The total gauge-fixed action $S_t$ is obtained from the high-temperature
action by adding to it the gauge-fixing action
\be
S_{gf} =\frac{1}{\tilde g_3^2(T)}
\int d^3x \frac{1}{2} {\cal F}_a {\cal F}_a
\ee
and the Fadeev-Popov action
\be
S_{FP} = \frac{1}{\tilde g_3^2(T)}
 \int d^3x \eta^\dagger (-\Delta + \frac{H_0^2}{4}) \eta \; .
\ee
It is the action $S_t= S_{ht}+S_{gf}+S_{FP}$
which has to be used in the definition of the
fluctuation operator (\ref{flucop}).

The Hilbert space of fluctuations decomposes into subspaces
defined by the symmetries of the background field. The
fluctuation operators given below have been derived
from those of Ref. \cite{BaaJu}. This analysis was
based on a $K$ spin basis ($\vec K = \vec J + \vec I$).
Alternatively  one might have used here
simply an analysis based on ordinary spin,
i.e. an expansion where
the Higgs field, the Fadeev Popov field
and the time components of the gauge fields are
expanded with respect to spherical harmonics and
the space components of the gauge fields with respect to
vector spherical harmonics $\hat x Y_l^m, r \nabla Y_l^m$ and
$\vec L Y_l^m$.

The electric components of the gauge field and the
isovector (would-be Goldstone) components of the Higgs field
form a coupled $(3 \times 3)$ system. The fluctuation
operator can be written in the
form $\bfm = \bfm^0 + {\bf V}$.
The free operator $\bfm^0$ is diagonal. It consists of
free partial-wave
Klein-Gordon operators
\be
\bfm^0 =-\frac{d^2}{dr^2}
-\frac{2}{r} \frac{d}{dr} + \frac{l_n(l_n+1)}{r^2} + m_n^2
\ee
 with masses $m_n$ given by $(0 ,0, m_H)$ respectively
 for the three components and with
centrifugal barriers corresponding to
angular momenta $l_n$ given analoguously by
 $(l+1,l-1,l)$. The nonvanishing components of the
potential are
\bea
V_{11}&=&V_{22}= H_0^2/4 \nonumber \\
V_{33}&=& H_0^2/4 + (\lambda_T/4 g^2)
(4 H_0^2-3 \epsilon H_0) \nonumber \\
V_{13}&=&V_{31}= -\sqrt{\frac{l+1}{2l+1}}\frac{dH_0}{dr}  \\
V_{23}&=&V_{32}= \sqrt{\frac{l}{2l+1}}\frac{dH_0}{dr} \; .
\eea

For $l=0$ the second component is absent due to the
vanishing of the vector spherical harmonic $ r \nabla Y_0^0$.
These amplitudes have a triple degeneracy due to
isospin besides the ordinary degeneracy $(2 l +1)$ from spin.

The fluctuation operator for the scalar part of the
Higgs field is given by
\bea \label{Higgflu}
\bfm& =& -\frac{d^2}{dr^2}
-\frac{2}{r} \frac{d}{dr} + \frac{l(l+1)}{r^2}
+ m_H^2  +V_{44}(r) \nonumber \\
 V_{44}&=& \frac{\lambda_T}{4 g^2} (12 H_0^2-6\epsilon H_0)
 \\
m_H^2 &=& \frac{\lambda_T}{4 g^2} (3\epsilon -4) \; .
\nonumber
\eea
This channel being an isosinglet its degeneracy is just $(2 l +1)$.

The time components of the gauge fields,
the Fadeev Popov fields and the magnetic components of the
vector potentials all satisfy the same equation
\be \label{EFP}
\bfm_l \psi_5 = \omega^2 \psi_5 \;.
\ee
It consists of a free massless partial wave Klein Gordon
operator
\be
\bfm^0 =-\frac{d^2}{dr^2}
-\frac{2}{r} \frac{d}{dr} + \frac{l(l+1)}{r^2}
\ee
and a potential
\be
V_{55} =\frac{H_0^2}{4}
\ee
which vanishes exponentially as $r \to \infty$.
There is no $l=0$ component of the magnetic vector potential
since the vector spherical harmonic
$\vec L Y_l^m$ vanishes. In the fluctuation determinant
all of these contributions cancel, only the
s-wave Fadeev-Popov contribution survives, due to the lack of
its magnetic counterpart. It is triply degenerate due
to isospin and has to be subtracted.

The partial-wave decomposition of the fluctuation operator
decomposes also its determinant,
\be
{\cal J} = \sum (2l+1) {\cal J}_l \; .
\ee
We now need a method for computing numerically the
determinants of the partial wave fluctuation operators.
Such a method
has been developed recently by V. G. Kiselev and the author
\cite{BaaKi} and will be presented briefly in the following
section.

\section{The fluctuation determinant of the \
electroweak bubble}
\setcounter{equation}{0}

A very fast method for computing fluctuation determinants
is based on a theorem on functional determinants;
references to earlier work and an elegant proof are given in
Ref. \cite{CoAS}.
Generalized to a coupled $(n \times n)$ system it can be
stated in the following way:

Let ${\bf f}(\nu,r)$ and ${\bf f}^0(\nu,r)$
denote the $(n \times n)$ matrices formed by
$n$ linearly independent solutions
$f_i^\alpha(\nu,r)$ and $f_i^{\alpha 0}(\nu,r)$ of
\be
(\bfm_{ij}+\nu^2) f_j^\alpha (\nu,r) =0
\ee
and
\be
( \bfm_{ij}^0 +\nu^2)
f^{\alpha 0}_j (\nu,r) =0  \; ,
\ee
respectively, with regular boundary conditions at $r=0$.
The lower index denotes the $n$ components, the
different solutions are labelled by the greek upper index.
Let these solutions be normalized such that
\be
\lim_{r \to 0} {\bf f}(\nu,r)({\bf f}^0(\nu,r))^{-1} = {\bf 1} \;.
\ee
Then the following equality holds:
\be \label{Flucdef}
{\cal J}(\nu) \equiv \frac{\det (\bfm +\nu^2)}{\det (\bfm^0+\nu^2)}
= \lim_{r \to \infty} \frac {\det {\bf f}(\nu,r)}{\det
{\bf f}^0(\nu,r)}
\ee
where the determinants on the left hand side are determinants
in functional space, those on the right
hand side are ordinary determinants of the $n \times n$ matrices
defined above.
If the theorem is applied at $\nu = 0$ it yields the desired ratio of
fluctuation determinants
${\cal J} \equiv {\cal J}(0)$. The consideration of
finite values of $\nu$ is
necessary in the discussion of zero modes.
The theorem has been applied for computing the 1-loop
effective action of a single scalar field on a bubble
background in Ref. \cite{BaaKi} and of a fermion
system on a similar background in Ref. \cite{BaaSoSu} previously.
It was found to yield very precise results, in addition
to providing a very fast computational method.

In the numerical application the solutions $f_n^\alpha$ were written
as \cite{Baa}
\be
f_n^\alpha(r) = (\delta_n^\alpha + h_n^\alpha(r))i_{l_n}(\kappa_n r)
\ee
with the boundary condition $h_n^\alpha(r) \to 0$ as $r \to 0$.
The values $l_n$ and $\kappa_n = \sqrt{m_n^2+\nu^2}$ depend on
the channel as specified in the previous section.
This way one generates a set of linearly independent
solutions which near $r=0$ behave like the free solution as required
by the theorem which then takes the form
\be
{\cal J}(\nu)
= \lim_{r \to \infty} \ln \det \{\delta_n^\alpha + h_n^\alpha(r)\}
\;.
\ee
The functions $h_n^\alpha(r)$ satisfy the differential
equation \cite{Baa}
\be
\frac{d^2}{dr^2}h_n^\alpha(r)+ \left ( \frac{2}{r}+ 2\kappa_n
\frac{i_{l_n}'(\kappa_n r)}{i_{l_n}(\kappa_n r)}\right)
\frac{d}{dr}h_n^\alpha(r)= V_{nn'}(r) \left( \delta_{n'}^\alpha
+ h_{n'}^\alpha(r)\right )\frac{i_{l_{n'}}(\kappa_{n'}r)}
{i_{l_n}(\kappa_n r)} \;
\ee
which can also easily be used for generating the functions
$h_n^\alpha$ order by order in $V$. In particular,
if this differential equation is truncated by leaving out the term
$h_{n'}^\alpha$ on the right hand side, one generates the
first order contribution to $h$ which is the tadpole term.
For more technical details we refer to Refs. \cite{BaaKi,BaaSoSu}.

With the partial wave fluctuation operators
given in the previous section the
application of the theorem to the case of the
electroweak bubble is straightforward.
Some points to be considered are \\
- the subtraction of the divergent
tadpole graphs \\
- double counting of gauge and would-be Goldstone fluctuations \\
- removing the translation zero mode \\
- removing a particular gauge zero mode. \\
We will discuss these briefly. We will add also some
remarks on details of the numerical computation.

\subsection{Tadpole diagrams}
The high-temperature three-dimensional theory has only
linear divergences of the form of tadpole diagrams
which renormalize the mass term of the Higgs field.
They have to be subtracted in the numerical
computation to obtain finite results. This was done in each partial
wave, for which the tadpole contribution may
be computed \cite{BaaKi,BaaSoSu} either by solving a
truncated differential equation or \cite{Baa} as an analytic
expression using the partial wave Green function. After these
contributions have been subtracted, the partial wave
contributions converge as $1/l^2$ and have a finite sum.

Of course this contribution has to be added back,
after having been regularized and renormalized.
Part of these diagrams have already been taken into account
in the renormalization of the four dimensional theory and
in giving the vacuum expectation value (\ref{scale1}) of the Higgs
field a quadratic temperature dependence. Some terms linear in
the temperature survive however and contribute
\cite{CarLi,BaaJu,ArMcL} (after dividing
by the temperature) to the 1-loop effective
action, i.e. the logarithm of the fluctuation determinant.
If the mass of the field in the
loop is $m_i$ and its coupling to the
external field
is described by the potential $V_i$
their contribution to the effective
action is given by $ -  m_i/8\pi \int d^3x V_i(r)$.
The fluctuating gauge fields have vanishing mass and do not
contribute. However we receive contributions from the
fluctuating Higgs fields. The mass circulating in the loop
is then $m_H$ which is - including the temperature dependence
and rescaling - given by equation (\ref{Higgflu}).
The potentials are
$V_{33}$ with triple isospin degeneracy and $V_{44}$.
So we have to restitute
the terms
\be \label{tad1}
S_{g-tad}^{1-l} =
- \frac{m_H}{2} \int dr r^2 \left(
\frac{3}{4}H_0^2 + 3\frac{\lambda_T}{4g^2} (4 H_0^2-
3 \epsilon  H_0)  \right )
\ee
for the gauge fields and
\be \label{tad2}
S_{h-tad}^{1-l} =
- \frac{m_H}{2} \int dr r^2
6\frac{\lambda_T}{4g^2} (2 H_0^2- \epsilon  H_0   )
\ee
for the Higgs field
to the 1-loop effective action.

We should like to remark on two slight
inconsistencies of this procedure.
The first one concerns our choice of the high temperature action.
We have adopted the action of Dine et al. \cite{DLHLL}
since it sets a certain standard and since it has been used also in
Ref. \cite{KriLaSch} to which we want to compare part of our results
(and which shares the inconsistency). In this action
the $T^2$ term does not include the contribution of the
Higgs loop tadpole. This can be seen from the coefficient
D in Eqs. (\ref{coeffs}) which should include a term
$m_H^2$ in addition to $3m_W^2+2 m_t^2$. The contribution was
neglected already in Ref. \cite{AnHa} ``taking the Higgs boson
sufficiently light''. Since the expression is dominated by the
top quark contribution whose fluctuations are not included at
all here this omission may be tolerated at the present level
of accuracy. In a more refined analysis it should and
can easily be remedied.

 The second point is the fact that we have
included already 1-loop effects into the tree level action, so
that part of our computation is now at the 2-loop level, without
constituting a complete and systematic 2-loop analysis. This applies
in particular also to the tadpole terms for which this could be
a more severe problem since they are the finite remnants of
divergent graphs. We can appeal here only to an argument - common
in many perturbative calculations - that possible inconsistencies
are of higher order and acceptable at an intermediate level
as they
will be cured in a complete higher order analysis.

\subsection{Double counting of gauge field fluctuations}

As mentioned in section 2 we are working with an action that
contains already the part of the 1-loop effective potential
induced by integrating out the gauge field and would-be
Goldstone boson fluctuations. These are present in the
temperature scale factors and couplings and appear especially
in the high temperature effective potential as the term
proportional to $\Phi^3$. While the $T^2$ contribution to
the vacuum expectation value (\ref{scale1})
comes from the tadpole diagrams and
has been taken into account along with these, the
$\Phi^3$ term is contained in our exact 1-loop effective
action. In order to avoid double counting it, this term has to be
subtracted from our numerical results. The incorporation
of this term into the tree level action was necessary in order
to obtain a first order phase transition and bubble
solution. If this was a good leading order approximation
the gauge field action should be well approximated by this
term. This is indeed the case (see below) but this implies also
that the remaining gauge and would-be Goldstone field contributions
are small differences of large terms, and that they cannot therefore
be expected to be very precise.

\subsection{Translation zero mode and unstable mode}
Translation invariance is broken by the classical solution,
so a zero mode appears. It occurs in the $l=1$ partial
wave of the fluctuation operator of the isoscalar part
of the Higgs field. It is easily removed using the
prescription given in \cite{BaaKi}: one applies the theorem
mentioned above at finite $\nu$ and defines
\be
{\cal J}_{l=1,Higgs} = \lim_{\nu \to 0} \lim_{r \to \infty}
\ln \left (\frac{\psi_4(\nu,r)}{\nu^2 i_l(\kappa r)}\right )
\ee
where $\kappa=\sqrt{\nu^2+m_H^2}$. Removing three
eigenvalues $\omega^2 =0 $ gives the fluctuation determinant
${\cal J}$ the dimension $ (energy)^{-6}$. The rate gets then a
dimension $(energy)^3 = 1 /(length)^3$. An additional dimension
$energy=1/time$ comes from the unstable mode prefactor
(see (\ref{rate})).The numerical
computation is based on energy units $ g \tilde v(T)$ which
are given in the Tables below.

The unstable mode makes the determinant of the p-wave
contribution negative. Replacing it by its absolute value
means just to revert the sign of the
determinant before taking the logarithm. We note that,
in contrast to Ref. \cite{KriLaSch} and in analogy to
Refs. \cite{CarLi,BaaJu}, we do not remove the zero mode
from the fluctuation determinant.

\subsection{Gauge zero mode}
Though we have imposed a gauge condition there is one residual
gauge degree of freedom. It is analoguous
to a constant
gauge function for the free theory. Indeed in the latter case
a constant
gauge potential $\Lambda(\vec x)= g_0$ does not contribute to the
vector potential and is therefore not eliminated by the
gauge condition $\partial_\mu a^\mu = 0$. In the case of the
bubble background field there is a similar {\it but nontrivial}
mode which satisfies the background gauge condition and is
therefore not eliminated by it. It manifests itself as a
zero mode
in the electric system for $l=0$.
The form of this mode (and the fact that it is really an
exact zero mode) was found after
extended numerical experiments.
It is given by a gauge
function $g(r)$ which satisfies the same
differential equation as the electric and Fadeev Popov modes,
Eq. (\ref{EFP}), i.e.
\be \label{gmode}
g'' +\frac{2}{r} g'-\frac{H_0^2}{4} g =0 \; .
\ee
With regular boundary condition at $r=0$ $g(r)$
becomes constant as $r \to \infty$, in
analogy to the free case.
Then the functions
\bea \label{gmod}
\psi_1(r)&=& - 2 g'(r) \nonumber  \\
\psi_3(r)&=& H_0 g(r)
\eea
satisfy the coupled system for the electric modes at $l=0$
which is given explicitly by
\bea \label{hdgl}
\psi_1'' + \frac{2}{r} \psi_1' -\frac{2}{r^2} \psi_1
&=& \frac{H_0^2}{4} \psi_1  - H_0' \psi_3  \nonumber \\
\psi_3'' + \frac{2}{r} \psi_3' - m_H^2 \psi_3
&=& \frac{H_0^2}{4} \psi_3   +\frac{\lambda_T}{4g^2}
(4 H_0^2-3\epsilon H_0) \psi_3 - H_0' \psi_1 \; .
\eea
It can be checked easily that this gauge zero mode
satisfies the background gauge condition (\ref{backgauge})
 and is therefore
not eliminated by it.  Since this zero mode is
not due to a symmetry broken by the
classical solution as the translation mode
it cannot be handled in the usual way.
On the other hand we observe that precisely for
the s-wave the Fadeev-Popov contribution has survived; furthermore
to each Fadeev-Popov mode with finite energy, i. e. a solution
of
\be
\psi_5'' +\frac{2}{r}
\psi_5'-\frac{H_0^2}{4} \psi_5 =-\omega_\alpha^2 \psi_5 \;
\ee
there is a solution of the electric s-wave system
constructed exactly as that for the gauge zero mode, i.e.
Eq. (\ref{gmod}) with $g$ replaced by $\psi_5$. So there
is a cancellation of all electric modes of this type
with the corresponding Fadeev-Popov ones, except for the mode
with $\omega_\alpha^2 =0$. There {\it is} of course a solution
of the Fadeev-Popov equation at this energy, but it is
``singular'' at infinity, going to a constant there.
The corresponding mode in the electric system is normalizable,
however, since only its derivative is involved in $\psi_1$ and
its product with the exponentially decreasing function $H_0$ in
$\psi_3$. The cancellation between the s-wave electric
modes (\ref{gmod}) and the Fadeev-Popov ones can be extended
therefore to the zero mode if the boundary condition
at $r \to \infty$ for the latter ones is replaced by
$\psi_5'(r) \to 0$.  This can be done in analogy with the
procedure described in the previous section by computing the
fluctuation determinant of the Fadeev-Popov mode at finite
$\nu$ via
\be \label{FPbc}
{\cal J}_{l=0,FP}(\nu)=
\ln \lim_{r \to \infty}
\left (\frac{\psi'_5(\nu,r)}{i'_l(\nu r)}\right) \;.
\ee
Then the Fadeev-Popov system at $l=0$ exhibits a zero mode as well,
the limit
\be
\lim_{\nu \to 0}
({\cal J}_{l=0,el}(\nu)-{\cal J}_{l=0,FP}(\nu))
\ee
is finite and defines the s-wave part of the fluctuation determinant.
In this way the Fadeev-Popov term cancels
all unwanted  longitudinal electric modes for $l=0$ - including
the one with frequency zero. We note that the change of boundary
condition as $r \to \infty$ affects only the s-wave and
only for massless fields.
The definition (\ref{FPbc}) yields results
identical to the usual one
(\ref{Flucdef}) if $l \neq 0$ and/or the fields are
massive.

\subsection{Some numerical details}
The analysis was performed as described in previous publications
\cite{BaaKi,BaaSoSu}. Contributions of angular momenta up to
$l_{max}=30$ were computed numerically, the higher ones were included
by performing a power fit $A l^{-2}+B l^{-3} + C l^{-4}$
through the last $5$ computed contributions
and by adding a corresponding
sum from $l_{max}$ to $\infty$.
This was done already at lower values of $l$, treating
the highest included angular momentum as the actual value of $l_{max}$.
The resulting expressions were found to be independent of
$l$ within typically four
significant digits for $l > 20$.

A more subtle point is the extrapolation to $r=\infty$ implied
in Eq. (\ref{Flucdef}). In the
previous analyses \cite{BaaKi,BaaSoSu} the fields had finite mass and the
approach to $r=\infty$ was exponential. For the massless fields
the Bessel functions $i_l(\kappa r)$ are replaced by
$r^l/(2l+1)!!$ and the functions
$h_1^\alpha$ and $h_2^\alpha$
 approach their asymptotic value only as $h_\infty + const./r$.
The extrapolation was performed using this Ansatz.
An exception occurs in the electric p-wave system, where
$h_2^\alpha$ picks up a logarithmic dependence on r due to
the cross term with $h_3^\alpha(r)$
on the r.h.s. of Eq. (\ref{hdgl}) which decreases only
as $1/r^2$. However this logarithmic dependence being
strictly proportional to $\delta^\alpha_3+h^\alpha_3(\infty)$,
i.e. to the third row of the matrix,
it does not contribute in the determinant, as also observed
numerically.

The tadpole contributions were computed in two ways,
once by solving a truncated differential equation
as described in \cite{BaaKi,BaaSoSu} and performing the analoguous
extrapolation, and once as
an integral using the partial wave Green function. In comparing the
two results the extrapolation was found - for the tadpole
contributions - to be reliable to four
significant digits typically.

Judging the accuracy of the results from the stability with
respect to varying extrapolations as $r\to\infty$ and
for large $l$ we would think that the purely numerical
part is accurate to
1 \%. The restituted tadpole contributions are given by
the expressions (\ref{tad1}) and (\ref{tad2}) whose evaluation
implies simple numerical integrals, they
can be considered as exact analytic expressions. This
restitution implies no delicate cancellations.
However even with a precision of 1\% for the numerical results
the final values of the gauge field
contribution have substantial errors since the
numerical part plus the tadpole contribution
is almost cancelled by the analytic
$\Phi^3$ contribution. Unfortunately, in contrast
to the sphaleron computation
\cite{BaaJu}, the cancellation ist not merely one
between two analytic expressions - the tadpole
and $\Phi^3$ contributions - but between the numerical results and
the analytic $\Phi^3$ contribution.

\section{Results and conclusions}
\setcounter{equation}{0}
The numerical results are given in Tables 1 to 4.
Here Table 1 is based on the values Higgs and
gauge boson masses $m_H=m_W=80.2$ GeV,
and a value of the top mass of $m_t= 170$ GeV. For the
vacuum expectation of the Higgs field we used
$v_0 =246$ GeV and for the gauge coupling the value
$g =.6516$. For the computation of Table 2 the values
$m_H=60$ GeV, $m_t=170$ GeV were used. Table 3 corresponds
to values $m_H =60$ GeV and $m_t= 140$ GeV, Table 4
to values $m_H=80.2$ GeV and $m_t= 140$ Gev; these
latter Tables are presented in order to compare with
results obtained in Ref. \cite{KriLaSch}
using the heat kernel expansion.
The values for the temperature chosen correspond to
10 equidistant steps of the quantity $\epsilon(T)$, defined
in Eq. (\ref{epsdef}), between the
onset of bubble nucleation at $\epsilon=2$ and
the critical temperature $T_0$ where bubble nucleation
ends at $\epsilon=4/3$. This choice is equivalent
to the choice of Kripfganz et al. \cite{KriLaSch} who
parametrize this range of temperatures by a variable
$y$ taking values between $0$ and $1$. Since Kripfganz et al. use
a somewhat different effective potential, the relation
between $y$ and $\epsilon$ is not precise, it is
essentially given by $y = 3-2\epsilon $ which we use as a
definition of `our' $y$. At small $y$ the
bubbles are large with thin walls, for $y \simeq 1$ the bubbles
are small and have thick walls.

Tables 1 and 2 are split into a part `a' which contains the essential
parameters for the minimal bubble and, in the last
column, the nucleation rate $R$ {\it without} fluctuation
corrections. The part `b' contains the fluctuation corrections,
i.e. the 1-loop effective action.
The results for $m_t=170$ GeV are given separately for the
isoscalar part of the Higgs field as $S^{1-l}_h$
and for the system of would-be
Goldstone fields and gauge fields (`gauge field contribution'
for short) as
$S^{1-l}_g$,
respectively. We give also separately the parts which were obtained
by the numerical analysis described in section 4, denoted as
$S^{1-l}_{h-num}$ and $S^{1-l}_{g-num}$, respectively.
The difference between
$S^{1-l}_{h}$ and $S^{1-l}_{h-num}$ is the tadpole contibution
$S^{1-l}_{h-tad}$ of Eq. (\ref{tad2}), and analoguously for the
gauge field. Note that the tadpole contributions to the Higgs
field are substantial.
The gauge field contribution $S_g^{1-l}$
contains the $\Phi^3$ part discussed in the previous section.
The numerical value of this term is given in the column
`$\Phi^3$'. This term should be close to the gauge field contribution,
and it is indeed. So the basic action used for computing the
bubble profiles represents a reasonable approximation to the
exact 1-loop effective action. The gauge field contribution
has to be reduced by this term since it would be double-counted
otherwise. The net gauge field contribution is denoted as
$\Delta S^{1-l}_g$ and given in the last column.
The correction to the rate can be simply obtained as a factor
$\exp (-\Delta S_{eff}^{1-l})$ where
$\Delta S_{eff}^{1-l} = S_h^{1-l} + \Delta S_g^{1-l}$.
The dimension
$energy^3$, here in units of $g \tilde v$, is already included
in the minimal bubble rate $R$.
One sees that the fluctuations  lead to a substantial
suppression of the nucleation rate.

While the result that the effective action for the
gauge fields is well approximated by the effective
potential, i. e. the $\Phi^3$ term, is very rewarding
a less comfortable feature appears
if one compares the 1-loop effective action with the tree level
action $\tilde S$. If the saddle point approximation which forms
the basis of transition rate formula (\ref{rate}) is justified,
then the 1-loop action should be smaller than the tree level one.
This is not the case. Large 1-loop corrections were found already by
Kripfganz et al. \cite{KriLaSch} when computing the
1-loop effective action for the Higgs-field fluctuation only.
We compare our results to theirs in Tables 3 and 4. Since
these authors define the fluctuation determinant differently
- they remove the unstable mode - we give, besides our
result $S^{1-l}_{Higgs}$, the expression $\ln (A/T^4)$ where
$A$ is the square root ${\cal J}^{-1/2}$ of the
fluctuation determinant with translation {\it and}
unstable modes removed.
The results are close to each other for small $\epsilon$ or
$y \simeq 1$, i .e. for small thick-wall bubbles.
For small $y$, i.e. for large thin-wall bubbles,
our exact results are systematically larger than the
approximate ones of Ref. \cite{KriLaSch}. The question of
finding reliable analytic estimates is certainly an interesting
one, especially the order in which the terms of the
heat kernel expansion are summed. In \cite{BaaSoSu} it was
found that a summation by the number of derivatives
(``derivative expansion'') yields very
precise results if the mass of the fluctuation is much
larger than the inverse size of the background field
configuration. In Ref. \cite{KriLaSch} the terms are
summed with respect to powers of the heat kernel time. The
deviation at small $y$ could be due to the fact that large
bubbles with thin walls have a very substantial derivative
contribution.
It will be interesting to make a more systematic analysis of
various analytical approaches.
The comparison of the Higgs effective action with the
leading minimal bubble action is less favorable than
found in Ref. \cite{KriLaSch}.
This is even more the case if the gauge loops are included, as one
sees from the previous Tables 1 and 2.

In conclusion we state three essential features of our results:
\\ - The 1-loop effective action is substantial, of the order of
and larger than the leading order minimal bubble action. This sheds
some doubt on the applicability of the semiclassical transition
rate theory.
\\ - The sign of the 1-loop effective action is
such that the transition rate is suppressed.
\\ - The 1 -loop ``$\Phi^3$'' contribution which has been
incorporated into the basic effective potential
is reproduced rather well by the 1-loop action. This means that this
term in the effective {\it potential} describes relevant
features of the effective {\it action}.

It will be interesting to pursue this subject further; it could be
of interest to try a selfconsistent extremalization of the sum of
leading and 1-loop action. This would certainly reduce the
total suppression. Furthermore it will be interesting
to see how the inclusion  of the fermion determinant
affects the transition rate.

\newpage
\section*{Table Captions}

{\bf Table 1a} Parameters of the minimal bubbles
for
$m_H = 60$ GeV and $m_t = 170$ GeV. The results are given as
a function of temperature in equidistant steps of the
variable $\epsilon$ (Eq. (\ref{epsdef})).
$\tilde v(T)$ is the temperature dependent vacuum expectation
value of Eq. (\ref{scale1}), $\lambda_T$ the
temperature dependent renormalized $\Phi^4$ coupling.
$\tilde S$ is the
minimal bubble action (or energy divided by $T$). $\omega_-^2$
is the square of the frequency of the unstable mode, given in units
of $g \tilde v (T)^2$.
The last column contains the logarithm of the nucleation rate
{\it without} the 1-loop corrections.
\\ \\
{\bf Table 1b}
$S_{h-num}^{1-l}$ is the 1-loop of the isoscalar part of the
Higs field as obtained in the numerical analysis.
$S_{h}^{1-l}$ is the total Higgs part of the 1-loop effective action,
obtained from $S_{h-num}^{1-l}$ by adding the tadpole contribution
$S_{h-tad}^{1-l}$.
 $S_{g-num}^{1-l}$ is the 1-loop
gauge and would-be Goldstone field action obtained by the
numerical analysis,
$S_g^{1-l}$ is again obtained by including the tadpole contribution.
The next colum gives the $\Phi^3$ term as included into the
high temperature action.
$\Delta S^{1-l}_g$ is the gauge field action after subtraction of
this $\Phi^3$ contribution. $\Delta S^{1-l}_{eff}$ is the total
effective action after removing the $\Phi^3$ contribution.
\\ \\
{\bf Table 2a} The same as Table 1a for $m_H = m_W = 80.2$ GeV and
$m_t = 170$ GeV.
\\ \\
{\bf Table 2b} The same as Table 1b for $m_H = m_W =80.2$ GeV
and $m_t = 170$ GeV.
\\ \\
{\bf Table 3} Comparison of the Higgs field effective action
with approximate results by Kripf\-ganz et al. for
$m_H =60$ GeV amd $m_t = 140$ GeV. The first entries are as defined
in the previous Tables. The quantity $A$ is the Higgs part
of the fluctuation determinant with removed unstable mode.
Our results are compared to the one of Ref. \cite{KriLaSch},
marked with the subscript $KLS$.
\\ \\
{\bf Table 4} The same as Table 3 for $m_H=80.2$ GeV
and $m_t=140$ GeV.

\newpage


\newpage
\section*{Tables}
\begin{center}
\vspace*{20mm}

\begin{tabular}{|c|c|c|c|c|c|c|c|}
\hline &&&&&&& \\
T [GeV] &$\epsilon$&$y$&$\tilde v$[GeV]&
$\lambda_T\times 10^2$&$\tilde S$&$\omega^2_-\times 10^3$
&$\ln (R[\rm GeV^4])$
\\ &&&&&&& \\
\hline &&&&&&&
\\
94.557&1.933 &.1&48.82&3.309& 1114.2&-0.1947&-1098.7\\
94.529&1.866 &.2&50.53&3.310&278.47 &-0.7911&-264.21\\
94.495&1.800 &.3&52.37&3.311&121.40&-1.819 &-107.83\\
94.455&1.733 &.4&54.33&3.312& 65.218&-3.338&-52.13 \\
94.405&1.663 &.5&56.47&3.314& 37.875&-5.438&-25.20 \\
94.347&1.600 &.6&58.72&3.316& 22.686&-7.814&-10.45 \\
94.276&1.533 &.7&61.19&3.319& 12.891&-9.760&-1.223 \\
94.191&1.466 &.8&63.85&3.322& 6.3659&-9.928&4.423 \\
94.089&1.400 &.9&66.75&3.325& 2.1002&-7.007&7.028 \\
&&&&&&& \\ \hline
\end{tabular}

\vspace*{10mm}
{\bf \large Table 1a}

\vspace*{20mm}
\begin{tabular}{|c|c|c||c|c|c|c||c|} \hline &&&&&&& \\
$y$&$S^{1-l}_{h-num}$&$S^{1-l}_{h}$
&$S^{1-l}_{g-num}$&$S^{1-l}_{g}$&$\Phi^3$&$\Delta S^{1-l}_{gauge}$
&$\Delta S_{eff}^{1-l}$
\\ &&&&&&& \\
\hline &&&&&&&
\\
.1&-389.2&505.0 &-85974 &-102494&-104089&1595&2100\\
.2&-81.21&133.6 &-8457  &-9952  &-10170 &218&351.6\\
.3&-28.77&64.25 &-1984  &-2285.5&-2338.3&52.83&117.1\\
.4&-11.78&28.30 &-663.6 &-739.33&-757.91&18.59&46.89\\
.5&-4.41 &28.85 &-263.44&-277.53&-287.56&10.03&38.88\\
.6&-7.54 &23.04 &-118.53&-113.26&-121.51&8.25&31.29\\
.7&1.48  &19.64 &-54.93&-42.63&-51.13&8.50&28.14\\
.8&3.11  &17.71 &-24.85&-10.04&-19.51&9.47&27.18\\
.9&4.79  &16.96 &-10.04&  5.53&-5.14 &10.67&27.63\\
&&&&&&& \\ \hline
\end{tabular}

\vspace*{10mm}
{\bf \large Table 1b}

\newpage
\vspace*{20mm}
\begin{tabular}{|c|c|c|c|c|c|c|c|} \hline &&&&&&& \\
T [GeV] &$\epsilon$&$y$&$\tilde v$[GeV]
&$\lambda_T\times 10^2$&$\tilde S$&$\omega^2_-\times 10^3$&
$\ln(R[{\rm GeV^4}]$)
\\ &&&&&&& \\
\hline &&&&&&&
 \\
115.725&1.933&.1&39.75&4.973&601.83&-0.294&-587.87\\
115.702&1.866&.2&41.16&4.974&151.04&-1.190&-138.32\\
115.675&1.800&.3&42.67&4.975&65.944&-2.733&-53.91\\
115.642&1.733&.4&44.29&4.976&35.314&-5.030&-23.76\\
115.602&1.663&.5&46.05&4.977&20.560&-8.173&- 9.42\\
115.555&1.600&.6&47.91&4.978&12.329&-11.73&- 1.61\\
115.498&1.533&.7&49.96&4.980& 7.066&-14.65& 3.09\\
115.428&1.466&.8&52.19&4.982& 3.450&-14.88& 5.82\\
115.345&1.400&.9&54.49&4.984& 1.145&-10.51& 6.64\\
&&&&&&& \\ \hline
\end{tabular}

\vspace*{10mm}
{\bf \large Table 2a}

\vspace{10mm}

\begin{tabular}{|c|c|c||c|c|c|c||c|} \hline &&&&&&& \\
$y$&$S^{1-l}_{h-num}$&$S^{1-l}_h$&
$S^{1-l}_{g-num}$&$S^{1-l}_g$&$\Phi^3$&
$\Delta S^{1-l}_{gauge}$&$\Delta S^{1-l}_{eff}$
 \\ &&&&&&& \\
\hline &&&&&&&
 \\
.1&-389.27&501.83& -44642& -54007&-56066&2060&2561\\
.2& -81.76&132.93&-4463.4&-5236.8&-5513.5&276.7&409.6\\
.3& -29.40&63.66&-1067.5&-1194.4&-1270.6&76.21&139.9\\
.4& -12.33&39.39&-362.64&-377.30&-409.36&32.06&71.45\\
.5&  -5.01&28.22&-148.57&-137.31&-155.97&18.66&46.88\\
.6&  -1.36&22.42&-69.30&-52.00&-66.02&14.04&36.46\\
.7&   0.87&19.03&-33.64&-15.43&-27.78&12.35&31.38\\
.8&   2.51&17.09&-16.35&  1.26&-10.56&11.82&28.91\\
.9&   4.20&16.36& -7.70&  8.91&- 2.80&11.71&28.07\\
&&&&&&& \\ \hline
\end{tabular}

\vspace*{10mm}
{\bf \large Table 2b}

\newpage
\vspace*{20mm}

\begin{tabular}{|c|c|c|c|c|c|c|c|} \hline &&&&&&& \\
T [GeV] &$\epsilon$&$\tilde S$&$\omega^2_-\times 10^3$&
$\tilde v$[GeV]&$ S^{1-l}_{Higgs}$&
$\ln \left( \frac{A}{T^4}\right)$&
$\ln \left (\frac{A}{T^4}\right)_{KLS}$ \\ &&&&&&& \\
\hline &&&&&&& \\
97.311&1.933 &1336.1 &-0.1686&56.88&505.2&-513.40&-115.\\
97.268&1.866 &336.57 &-0.6965&58.87&134.2&-141.56&-82.0\\
97.216&1.800 &146.77 &-1.601 &61.00&64.60&-71.40&-50.6\\
97.153&1.733 &78.407 &-2.955 &63.30&40.20&-46.54&-36.7\\
97.078&1.663 &45.935 &-4.777 &65.75&29.13&-35.06&-29.5\\
96.988&1.600 &27.934 &-6.891 &68.38&23.23&-28.83&-25.4\\
96.879&1.533 &15.554 &-8.613 &71.24&19.83&-25.15&-23.0\\
96.748&1.466 &7.6707 &-8.758 &74.34&17.89&-23.01&-21.8\\
96.590&1.400 &2.5103 &-6.155 &77.70&17.18&-22.30&-22.0\\
&&&&&&& \\ \hline
\end{tabular}

\vspace*{10mm}
{\bf \large Table 3}

\vspace*{20mm}
\begin{tabular}{|c|c|c|c|c|c|c|c|} \hline
&&&&&&& \\
T & $\epsilon$&$\tilde S$&$\omega^2_-\times 10^3$&$ \tilde v$ [GeV]&
$ S^{1-l}_{Higgs}$&
$\ln \left(\frac{A}{T^4}\right)
$&$\ln \left(\frac{A}{T^4}\right)_{KLS}$\\ &&&&&&& \\
\hline &&&&&&&  \\
123.815&1.933&598.02&-0.296&43.12&490.0&-500.0&-89. \\
123.783&1.866&153.96&-1.175&44.63&132.83&-142.0&-81.4\\
123.744&1.800&67.333&-2.695&46.26&63.712&-72.32&-52.0\\
123.697&1.733&36.144&-4.949&48.01&39.485&-47.61&-38.3\\
123.640&1.666&21.072&-8.035&49.92&28.306&-36.06&-31.0\\
123.572&1.600&12.563&-11.58&51.95&22.428&-29.84&-26.8\\
123.490&1.533&7.1533&-14.44&54.16&19.047&-26.18&-24.4\\
123.391&1.466&3.5336&-14.67&56.57&17.118&-24.06&-23.1\\
123.270&1.400&1.1142&-10.25&59.22&16.363&-23.30&-23.3\\
&&&&&&&  \\
\hline
\end{tabular}

\vspace*{10mm}
{\bf \large Table 4}
\end{center}

\end{document}